\documentclass[aps,twocolumn,superscriptaddress,nofootinbib,floatfix]{revtex4-2}

\usepackage{amsmath}
\usepackage{mathrsfs}
\usepackage{amssymb}
\usepackage{bm} 
\usepackage{graphicx,grffile,textcomp}
\usepackage{dcolumn} 
\usepackage[usenames,dvipsnames]{xcolor}
\usepackage{multirow}
\usepackage{epstopdf}
\usepackage{mathtools} 
\usepackage{tabularx}
\usepackage{microtype}
\usepackage{comment}
\usepackage[version=4]{mhchem} 
\usepackage{array}
\usepackage{float}
\usepackage{wrapfig} 
\usepackage{url}
\usepackage{hyperref}
\usepackage{cleveref}
\usepackage[utf8]{inputenc}
\usepackage[T1]{fontenc}
\usepackage{bbold}

\hypersetup{colorlinks=true, linkcolor=blue, urlcolor=blue, citecolor=blue}

\definecolor{myred}{RGB}{230, 7, 77}

\definecolor{blue(ryb)}{rgb}{0.01, 0.28, 1.0}
\definecolor{jade}{rgb}{0.0, 0.66, 0.42}
\definecolor{cadmiumred}{rgb}{0.89, 0.0, 0.13}
\definecolor{darkviolet}{rgb}{0.58, 0.0, 0.83}

\begin{document}

\title{Electrostatic effects in nano-reactor-confined charge regulated macroions}

\author{Manit Klawtanong}
\affiliation{Department of Physics, Faculty of Science, Ramkhamhaeng University, Bang Kapi, 10240 Bangkok, Thailand}

\author{Petch Khunpetch}
\email{petch.k@rumail.ru.ac.th}
\affiliation{Department of Physics, Faculty of Science, Ramkhamhaeng University, Bang Kapi, 10240 Bangkok, Thailand}

\author{Huaqiong Li}
\affiliation{
Zhejiang Key Laboratory of Soft Matter Biomedical Materials, Wenzhou Institute of the University of Chinese Academy of Sciences, Wenzhou, Zhejiang 325000, China}

\author{Shigeyuki Komura}
\affiliation{
Zhejiang Key Laboratory of Soft Matter Biomedical Materials, Wenzhou Institute of the University of Chinese Academy of Sciences, Wenzhou, Zhejiang 325000, China}





\begin{abstract}
We formulate a thermodynamic model of a nano-reactor containing charge-regulated macroions within an electrolyte-permeable enclosure. The model is then formalized within the Poisson-Boltzmann electrostatics augmented by the consistent inclusion of the charge dissociation of molecular groups residing on the surface of the entrapped macroions via charge regulation formalism. By solving the basic equilibrium equations in the linearized Debye-H\"uckel type approximation, we analyze the salient features of the inhomogeneous electrolyte distribution and macroion charge. We found that the surface charge asymmetry/symmetry of the macroions strongly affects the spatial profile of electrostatic potential. The effective screening length shows the non-monotonic behavior, arising from the complex interplay between the bathing external solution and macroion effective charges, which govern charge regulation equilibria. The total pressure at the nano-reactor enclosure boundary decreases monotonically as the enclosure radius and the ionic bulk salt concentration increase. Also, the resulting pressure is strongly influenced by the surface charge densities of the nano-reactor and the number of confined macroions.
\end{abstract}

\maketitle

\section{\label{sec:introduction}Introduction}
Functional hollow colloidal nanospheres show great promise in many nano-reactor and nanocarrier technologies and have spawned different new research directions to understand their properties.~\cite{LEE2014631} Among these nano-reactor technologies, one could list self-assembled, thermostable, ferritin protein nanocages,~\cite{doi:10.1021/acsbiomedchemau.2c00003} lipid non-viral vector nanoparticles as a delivery platform for nucleic acids,~\cite{Inguva2024} colloidal carbon sphere nano-reactors for energy storage, electrochemical conversion, and catalysis,~\cite{adma.201903886} surfactant micellar nano-reactors in green chemistry,~\cite{SORHIE2022100875} enzyme encapsulation in protein nanocontainers in order to study deviations from well-understood bulk processes,~\cite{acssynbio.5b00037} as well as many others. In order to understand the thermodynamic properties of nanoparticles and their functioning as nano-reactors, it is important to formulate models that could predict their stability, selectivity, and responsiveness.~\cite{renggli2011selective} Since most of the components of the hollow colloidal nanospheres are charged, the understanding of the role of electrostatics in thermodynamic equilibria is particularly important, if not outright essential.

Electrostatic interactions are a crucial component in controlling the assembly of nanostructured materials~\cite{Isr11,Muthu2023} and are one of the essential components among the salient interactions at the nanoscale.~\cite{RevModPhys.82.1887} While the general theory of electrostatic interactions in colloid electrolyte solutions, based on the Poisson-Boltzmann mean-field paradigm, is well established,~\cite{Saf18} the modifications in the simple interaction potential wrought about by the dissociation equilibria at colloidal interfaces are often overlooked.~\cite{avni2019charge} These amounts to assuming that either the electrostatic charge at the macromolecular surface or the electrostatic potential is constant.~\cite{Saf18}  Although this simplification makes the problem of electrostatic interactions more tractable, the naturally occurring nanoparticle and macromolecular surfaces of  interest, e.g., hard colloidal particles,~\cite{Jianzhong2020b} soft biological molecules including proteins,~\cite{Jianzhong2020} and lipid membrane vesicles,~\cite{Khunpetch2022,Khunpetch2023}  rarely satisfy this assumption. They respond to their environment by modifying the dissociation equilibrium that then modifies both the surface charge as well as the surface potential, adjusting them according to the separation between them and the bathing solution conditions.~\cite{Tre16,Tre17} This conceptual framework that consistently includes the charge dissociation equilibria into the general theory of electrostatic interactions is formally referred to as the \textit{charge regulation} (CR) paradigm.~\cite{Nin71,Avn19}  

Phospholipids and proteins are among the biological macromolecules that specifically exhibit charge dissociation equilibria. The charge of phospholipid polar heads originates in deprotonated phosphate  groups, protonated amine group, and deprotonated carboxylate group, with different  corresponding dissociation constants.
~\cite{Majee2016,Khunpetch2022,Khunpetch2023} The charge of proteins originates in dissociable amino acids, specifically from the deprotonated carboxylate on the side chains of aspartic and glutamic acid, 
the deprotonated hydroxyl of the phenyl group of tyrosine, 
and from the protonated amine group of arginine and lysine. 
The protonated state of the secondary amine of histidine 
could also contribute to the charge of the capsid.~\cite{Simonson_2003, Nap14,D1SM00232E} For a hollow proteinaceous or phospholipid shell encaging macroion cargo, such as other proteins and/or nucleic acids, the imprint of charge regulation will therefore be present in every aspect of electrostatics, adjusting the dissociation equilibria of the cargo to the dissociation equilibria of the container. 

Devising a tractable as well as a realistic framework that would allow for a solvable model system coupling the dissociation processes and electrostatic interactions is difficult.~\cite{Inguva2024,Schoot} We thus delimit ourselves to a simplified case that would retain some of the salient features of the full problem, specifically including the charge regulation of the molecular cargo in as complete a form as possible. We refer to this level of detail as the Poisson-Boltzmann charge-regulated (PB-CR) theory. To this effect, we consider a solution of fully charge regulated macroions confined within an electrolyte-permeable enclosure containing fixed surface charge densities on its outer and inner surfaces.
This model bears some resemblance to virus nanoparticles ~\cite{arul2024viral} and/or lipid nanoparticles,~\cite{Inguva2024} containing proteinaceous, enzymatic cargo. The number of macroions inside the enclosure is not set by the equilibrium with the bulk, but is rather a consequence of the formulation protocol, remaining constant for any environment condition. This effectively implies an enclosure that is rigid enough to be able to contain the macroions at any bathing solution conditions. By solving the linearized Debye-Hückel equation analytically, we found that the spatial electrostatic potential profile, and spatial density distribution of all the mobile ionic components are clearly affected by the surface charge symmetry/asymmetry of the entrapped macroions. Strongly influenced by the surface charge densities of the enclosure and the number of macroions, we found that the pressure at the nano-reactor surface monotonically decreases as the enclosure radius and the ionic bulk salt concentration increase. On the other hand, our results suggest the non-monotonic behavior of the effective screening length due to the interplay between the bathing external solution and macroion effective charges.

This paper is organized as follows. 
In Sec.~\ref{sec:developments}, we formulate the model based on the Poisson-Boltzmann mean-field theory and the charge regulation formalism. We then examine the linearized Debye-Hückel equation and introduce the total pressure acting on the boundary of the spherical enclosure. 
In Sec.~\ref{sec:results}, we present our numerical results and discuss the salient features of the model system. Finally, in Sec.~\ref{sec:cons}, we summarize our present work with brief remarks and suggestions for future work.

\section{\label{sec:developments}Model and formalism}

\begin{figure}[t]
\centering
\includegraphics[scale=0.75]{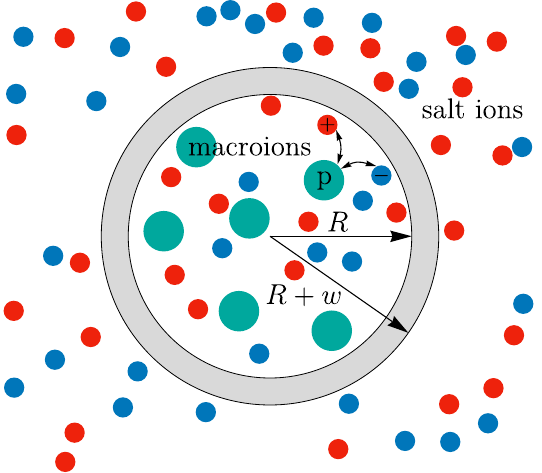}%
\caption{Schematic rendition of the molecular model. Electrolyte ions are exchangeable between the inside and the outside of the spherical enclosure (nano-reactor) and are thus in grand canonical equilibrium, characterized by a fixed chemical potential. The macroion cargo, on the other hand, is entrapped inside the enclosure and is thus not in grand canonical equilibrium with the bulk but can be seen as corresponding to a canonical subsystem with a fixed number of macroions, set by the experimental (preparation) procedure. Charge regulation of the macroions is indicated by the arrows indicating the adsorption/desorption charge regulation equilibrium. The inner and outer surfaces of the shell are assumed to carry fixed surface charge densities $\sigma_R$ and $\sigma_{R+w}$ of constant magnitude. The dielectric constant of water is $\varepsilon_w$, and the dielectric constant of the shell is $\varepsilon_m$.}
\label{schematic}
\end{figure}

We consider a hollow proteinaceous and/or phospholipid nano-reactor shell with a uni-valent electrolyte solution that can equilibrate with the bulk solution, 
and a fixed number of macroions with CR sites, entrapped inside the nano-reactor and cannot equilibrate with the bulk, as shown in Fig.~\ref{schematic}. We assume that the nano-reactor is a spherical enclosure with inner radius $R$ and finite thickness $w$. Additionally, the inner and outer surfaces of the shell are assumed to carry fixed surface charge densities $\sigma_{R}$ and $\sigma_{R+w}$, respectively. 
The macroions are assumed to have the CR sites that can adsorb/desorb salt ions from/to the solution. 
The details of the charge regulation and electrostatics are discussed in Ref.~\cite{Avni18} and we follow an analogous formulation, taking into account the details of our model setup. Clearly, the model system can be identified as a grand canonical ensemble for the electrolyte ions, and a canonical ensemble for the confined macroion cargo.

In our model, the macroions are assumed to have two types of active sites. One can adsorb/desorb a cation from/to the solution, and the other one can adsorb/desorb an anion. This adsorption model can be described by the chemical reactions:
\begin{eqnarray*}
    \text{A} + \text{B}^{+} &\rightleftarrows& \text{AB}^{+}\nonumber\\
    \text{C} + \text{D}^{-} &\rightleftarrows& \text{CD}^{-},\nonumber\\
\end{eqnarray*}where A (C) is the active site that can adsorb/desorb $\text{B}^{+} (\text{D}^{-})$ from/to the salt solution. 
We further assume that each macroion contains $N_{+}$ potentially dissociable $\text{A}$ sites
and $N_{-}$ dissociable 
$\text{C}$ sites, and only a fraction $\phi_{+} (\phi_{-})$ of the overall $N_{+} (N_{-})$ sites will become charged at any given solution condition. Moreover, the A and C sites are assumed to be uncorrelated in any direct manner.

The total free energy of the system composed of the electrostatic component, the mixing entropy, and the non-electrostatic CR free energy of the macroion cargo can be written as
\begin{eqnarray}
\label{bp1}
F\left[n_{\pm},n_p,\psi,\phi_{\pm}\right] &=& \int_V\text{d}^{3}r\Bigg[-\frac{\varepsilon_0 \varepsilon_w}{2}\left(\nabla\psi\right)^{2}\nonumber\\
&& +\left(en_{+}-en_{-} +  n_p \sum_{i=\pm}N_{i}\phi_{i} e_{i}\right)\psi\nonumber\\
&& - TS +\, n_p  g(\phi_{\pm}) -  \left(\mu_{+}n_{+}  +  \mu_{-}n_{-} \right)\Bigg],\nonumber\\
\label{freenergy}
\end{eqnarray}
where the volume integral extends over the region inside the enclosure, $\varepsilon_0$ and $\varepsilon_w$ are the vacuum permittivity and the dielectric permittivity of the aqueous bathing solution, respectively, $e_{\pm}=\pm e$ ($e > 0$) are the charges of the monovalent salt solution, $T$ is the temperature and $S$ is the entropy. The local electrostatic potential is denoted by $\psi$, and the chemical potentials of the electrolyte ions are $\mu_{\pm}$, 
$g(\phi_{\pm})$ is the single macroion free energy characterizing the specific CR process, and is assumed to be described by the Langmuir isotherm:~\cite{Avni18}
\begin{eqnarray}
g(\phi_{\pm}) &=& -\sum_{i=\pm}N_{i}\phi_{i}\left( \mu_{i}+\alpha_{i}\right)\nonumber\\
&&\, +\, k_{B}T\sum_{i=\pm} N_{i}\left[\phi_{i}\ln \phi_{i} +(1-\phi_{i})\ln(1-\phi_{i})\right],\nonumber\\
\label{CRenergy}
\end{eqnarray}
where $\alpha_{+} (\alpha_{-})$ is the free-energy change in adsorbing/desorbing a cation (anion). If $\alpha_\pm>0$, it means that there is a free-energy gain in association, while $\alpha_\pm < 0$ opposes such association and promotes dissociation.


In the dilute limit of both the electrolyte ions and macroions, the entropy can be approximated by the ideal mixing entropy
\begin{equation}
    S = -k_B\sum_{i=\pm}n_{i}\left[\ln(n_{i}a^{3})-1\right]
 - k_B n_p\left[\ln(n_p\gamma a^{3})-1\right],
\end{equation}
where, $n_{\pm}$ and $n_{p}$ are the number densities of the electrolyte cations/anions and macroions, respectively, and $k_{B}$ is the Boltzmann constant. Here, we assume that the electrolyte ions and the solvent molecules have the same volume, $a^3$, and the specific volume of the macroion is $\gamma a^3$, where the numerical prefactor, $\gamma > 1$, describes the ratio between the two molecular volumes.

Note also that in what follows we will assume a spherical symmetry of the thermodynamic equilibrium state so that all the fields entering the free energy depend only on the radial distance, $\vert{\bf r}\vert = r$.

\subsection{The PB-CR equations}
Minimizing the free energy functional $F[n_{\pm},n_p,\psi,\phi_{\pm}]$, Eq.~(\ref{freenergy}), with respect to all the variables yields the thermodynamic equilibrium conditions. We obtain the Poisson-Boltzmann charge regulation (PB-CR) equations that combine the mean-field electrostatics with the variable charge of the mobile macroions.~\cite{Saf21} 

Minimization of $F[n_{\pm}, n_p, \psi, \phi_{\pm}]$ with respect to the density variables $n_{\pm}$ and $n_p$, $\delta F/\delta n_{\pm}=0$ and $\delta F/\delta n_{p}=0$, yields the following equations 
\begin{equation}
\pm\beta e \psi + \ln(n_{\pm}a^{3}) - \beta\mu_{\pm} = 0,
\label{npm}
\end{equation}
and
\begin{equation}
\beta \sum_{i=\pm}N_{i}\phi_{i} e_{i} \psi + \beta g(\phi_{\pm}) + \ln(n_{p}\gamma a^{3}) = 0,
\end{equation}
respectively, with $\beta = 1/(k_B T)$. From Eq.~(\ref{npm}), it further follows that
\begin{eqnarray}
n_{\pm}(\psi)=n^\pm_{b}{\rm e}^{\mp \beta e\psi},
\end{eqnarray}
where $n^{\pm}_{b}= [\exp(\beta\mu_\pm)]/a^3$ is defined as the cation/anion bulk concentration, taken at zero reference potential, $\psi=0$, while
\begin{equation}
n_p(\psi) = n_p^{b}\,{\rm e}^{-  \beta \sum_{i=\pm}N_{i}\phi_{i} e_{i} \psi  - \beta\left[g(\phi_{\pm})- g_0\right]},
\label{eqnp}
\end{equation} where $g_0$ acts as a reference bulk state with $\psi=0$, and the macroion bulk density $n_p^{b}$ satisfies $n_p^{b} = [\exp (-\beta g_0)]/(\gamma a^3)$. Furthermore we will stipulate that the macroions are not in equilibrium with a bulk reservoir as there are no bulk macroions, but are confined to the nano-reactor interior depending on the method of preparation. The number of macroions inside the nano-reactor is assumed as fixed and given by $N_p$, therefore
\begin{equation}
\int_V n_p(\psi(r))\, \text{d}^{3}r = N_p,
\label{const}
\end{equation} where, again, the integral extends over the region inside the nano-reactor. This is the main difference with the case of equilibrated macroions analyzed before~\cite{Avni18} since in our case the macroions are trapped within the shell and cannot exchange with the bathing solution outside the nano-reactor. In our calculation, we assume that $n_p^{b}$ is fixed, but is then chosen such that Eq.~(\ref{const}) is satisfied.

Minimization of $F[n_{\pm}, n_p, \psi, \phi_{\pm}]$ with respect to the potential $\psi$, $\delta F/\delta \psi =0$, yields a generalized Poisson-Boltzmann equation,
\begin{equation}
-\varepsilon_0 \varepsilon_w\nabla^2\psi = \rho(\psi)=
e\left[ n_{+}(\psi) - n_{-}(\psi)\right]+ Q(\psi) n_p(\psi),
\label{eqPB0}
\end{equation}
where $Q(\psi) = \sum_{i=\pm} \, N_{i}\phi_{i} e_{i}$ is the effective charge of the macroions. Note that if the macroions were simple ions of valency $\pm z$, the effective charge is $Q=\pm ze$. Consequently, $g$ in Eq.~(\ref{CRenergy}) has the usual form $\pm e z \psi$, and the above equations would reduce to the standard form of the Poisson-Boltzmann equation.~\cite{Mar21}

We finally minimize the free energy functional with respect to the occupied ionic fractions on the surface of the trapped macroions $\phi_\pm$, $\delta F/\delta\phi_\pm=0$. This yields the Langmuir-Davies isotherm~\cite{Andelman1996}
\begin{equation}
N_{\pm} e_{\pm} \psi + \frac{\partial g(\phi_{\pm})}{\partial \phi_\pm} = 0.
\end{equation}
Taking into account the form of the charge regulation free energy, Eq.~(\ref{CRenergy}), this can be further cast as
\begin{eqnarray}
 e_{\pm} \psi -  \left( \mu_{\pm}+\alpha_{\pm}\right) + k_BT \ln{\frac{\phi_{\pm}}{(1-\phi_{\pm})}} = 0,
\end{eqnarray}
the solution of which can be written in the following form
\begin{eqnarray}
\phi_{\pm} 
&\equiv& \frac{{Z}_{\pm}(\psi)-1}{Z_{\pm}(\psi)},
\label{equphi}
\end{eqnarray}
where we have introduced 
\begin{eqnarray}
Z_{\pm}(\psi) = 1+n_b ^{\pm}K_\pm \, {\rm e}^{- \beta e_{\pm} \psi},
\end{eqnarray}
the partition function of a single macroion, with chemical equilibrium constants $K_\pm=a^{3}{\rm e}^{\beta\alpha_\pm}$. We can now substitute $\phi_{\pm}$ into the CR free energy in Eq.~(\ref{freenergy}) and obtain
\begin{equation}
g(\phi_{\pm}) =  - \sum_{i=\pm} N_i \phi_i e_{i} \psi- k_{B}T \sum_{i=\pm} N_i \ln{Z_{i}(\psi)}.
\end{equation}
From here the macroion density Eq.~(\ref{eqnp}) can be  furthermore written as
\begin{equation}
n_p(\psi) = n_p^{b} \left(\frac{{Z_{+}(\psi)}}{{Z_{+}(0)}}\right)^{N_+} \left(\frac{{Z_{-}(\psi)}}{{Z_{-}(0)}}\right)^{N_-},
\label{nppsi}
\end{equation}
and the effective charge of the macroions can also be expressed via the partition function of the macroions as
\begin{equation}
Q(\psi) =e\left(N_+\frac{{Z}_{+}(\psi)-1}{Z_{+}(\psi)} - N_-  \frac{{Z}_{-}(\psi)-1}{Z_{-}(\psi)}\right).
\label{Qdef}
\end{equation}
The final form of the PB equation, that can be obtained by inserting the equilibrium equations satisfied by the ionic concentrations $n_{\pm}$, $n_p$, and $\phi_{\pm}$ into Eq.~(\ref{eqPB0}), then reads
\begin{eqnarray}
-\varepsilon_0 \varepsilon_w\nabla^2\psi &=& e\left( n^+_{b}{\rm e}^{- \beta e\psi} -  n^-_{b}{\rm e}^{\beta e\psi}\right)\nonumber\\
&&\,+\, en_p^{b} \left(\frac{{Z_{+}(\psi)}}{{Z_{+}(0)}}\right)^{N_+}\!\! \left(\frac{{Z_{-}(\psi)}}{{Z_{-}(0)}}\right)^{N_-}\nonumber\\
&&\quad\times \left( N_+ \frac{{Z}_{+}(\psi)-1}{Z_{+}(\psi)} - N_-   \frac{{Z}_{-}(\psi)-1}{Z_{-}(\psi)}\right).\nonumber\\
\label{PBCR1}
\end{eqnarray}
Note that this is the PB equation valid inside the spherical enclosure, i.e., for $r \leq R$. Within the shell, $R\leq r\leq R+w$, we have 
\begin{eqnarray}
-\varepsilon_0 \varepsilon_m\nabla^2\psi &=& 0,
\label{PBCR3}
\end{eqnarray}
with $\varepsilon_m$ being the dielectric constant of the shell. Outside, $r \geq R+w$, there are no macroions and the PB equation is simplified to its standard form, 
\begin{equation}
-\varepsilon_0 \varepsilon_w\nabla^2\psi 
= -2en_{0}\sinh\left(\beta e\psi\right),
\label{PBCR2}
\end{equation}
where $n_{0}$ is the bulk salt concentration. 

In addition, we assume that the system is at infinite dilution, meaning that there is only a single nano-reactor at the origin, while the bathing external solution contains only simple electrolyte. The charge electro-neutrality of the system, i.e., the integral of the total charge density, including the surface charge densities, stipulates that the boundary conditions at the inner and outer surfaces of the shell are in the form
\begin{equation}
    {\bf n}\cdot\left[ \varepsilon_0 \varepsilon_w \nabla \psi(R) - \varepsilon_0 {\varepsilon_m}\nabla \psi(R) \right] = \sigma_R
    \label{BC1}
\end{equation} 
and 
\begin{equation}
{\bf n}\cdot\left[ \varepsilon_0 {\varepsilon_m} \nabla \psi(R+w) - \varepsilon_0 \varepsilon_w \nabla \psi(R+w) \right]= \sigma_{R+w},
    \label{BC2}
\end{equation}
where $\bf n$ is the normal to the surface.


The above three equations, 
Eqs.~(\ref{PBCR1}), (\ref{PBCR3}), and (\ref{PBCR2}), together with the boundary conditions Eqs.~(\ref{BC1}) and (\ref{BC2}), and the condition Eq.~(\ref{const}),  completely define the thermodynamic equilibrium of the system and we refer to them as the PB-CR set of equations.

To the basic equilibrium equations, we can furthermore add the mechanical pressure acting inside and outside the shell because of the inhomogeneous distribution of the concentrations of the relevant species. The derivation of the pressure follows closely the analogous derivation for the simple PB case.~\cite{Mar21} In general, the pressure is obtained from the matrix elements 
of the pressure tensor which is defined as~\cite{lan84,Trizac97}
\begin{eqnarray}
P_{\alpha \beta} &=& P^E_{\alpha \beta} + P^O_{\alpha \beta},
\label{stresses}
\end{eqnarray}
where we have decomposed the pressure tensor into its {\sl electrostatic} (negative of the Maxwell stress tensor) component, 
\begin{equation}
P^{E}_{\alpha \beta} = \varepsilon_0\varepsilon_w\left(\frac{1}{2}(\nabla\psi)^2\delta_{\alpha \beta}-\nabla_{\alpha}\psi  \nabla_{\beta}\psi\right),
\end{equation}
and {\sl osmotic} (van't Hoff osmotic pressure difference) component,
\begin{equation}
P^O_{\alpha \beta} = k_{B}T\left[\sum_{k=\pm}\left(n_{k}\left(\psi\right)-n_{b}^{k}\right)+\left(n_{p}\left(\psi\right)-n_{p}^{b}\right)\right]\delta_{\alpha \beta}.
\label{eq:osmotic_force} 
\end{equation}
Note that here we have ion densities ($+, -$ and $p$) and not charge densities. The total pressure acting on the boundary of the enclosure is then given by the difference 
\begin{equation}
p = p_{\text{in}} - p_\text{out},  \label{forcetot}
\end{equation}
where $p_{\text{in}} = P_{rr}(r = R)$ and $p_{\text{out}} = P_{rr}(r = R + w)$ are the (bulk) pressures on the inside and outside surfaces, respectively. 


In what follows, we will not pursue further the full non-linear formulation of the problem but will instead concentrate on the solution of the linearized version. Previous related analyses of charge regulated problems have shown that the linearized version yields qualitatively and, often, quantitatively similar results as the non-linear analysis.~\cite{Khunpetch2022,Khunpetch2023}

\subsection{\label{sec:linearization}Linearized PB-CR equations}

The linearization follows an analogous route to the 
Debye-H\"uckel type approximation,~\cite{Mar21} with the additional proviso that one needs to expand also the CR quantities to the linear order in the electrostatic potential.~\cite{sunita2022,sunita2024} To the lowest, linear order in the electrostatic potential $\psi$ valid inside the spherical enclosure, $r \leq R$, we then obtain from Eq.~(\ref{eqPB0})
\begin{eqnarray}
\rho(\psi) &\simeq& \rho\left(0\right) - \varepsilon_{0}\varepsilon_{w}\kappa_{\text{eff}}^{2}\psi  +  {\cal O}\left(\psi^2\right), 
\end{eqnarray}
where $\kappa_{\text{eff}}$ is an inverse effective screening length and is given by 
\begin{align}
\kappa_{\text{eff}}^{2} &= -\frac{1}{\varepsilon_{0}\varepsilon_{w}}\left(\frac{\partial \rho\left(\psi\right)}{\partial\psi}\right)\Big|_{\psi =0} \nonumber\\
 &= \kappa_{0}^{2} \Bigg\{1+\frac{1}{2}\frac{ n_{p}^{b}}{n_{0}}\Bigg[\frac{(n_{b}^{+}+n_{b}^{-}-2n_{0})}{n_{p}^{b}}\nonumber\\
&\quad \,  +\frac{Q^{2}(0)+[\Delta Q(0)]^{2}}{e^{2}}\Bigg]\Bigg\},
 \label{eq:kappai01}
\end{align}
with
\begin{eqnarray}
 [\Delta Q(0)]^2 &\equiv& - k_BT \frac{\partial Q(\psi)}{\partial \psi}\Big|_{\psi = 0}.
\end{eqnarray}
Again, $Q(\psi)$ is defined in Eq.~(\ref{Qdef}) and $\kappa_{0}=\sqrt{2e^2 n_0/(\varepsilon_w\varepsilon_0 k_BT)}$ is the inverse Debye screening length. Furthermore we assume the electro-neutrality condition in the bulk, $\rho\left(0\right)=0$, so $n^{\pm}_{b}$  obey~\cite{Avni18}
\begin{eqnarray}
n^{\pm}_{b} = n_{0}-n_{p}^{b}N_{\pm}\phi_{\pm}\left(0\right).
\end{eqnarray}
Taking into account the electro-neutrality condition, the inverse effective screening length $\kappa_{\text{eff}}$ in Eq.~(\ref{eq:kappai01}), can be written as 
\begin{eqnarray}
\kappa_{\text{eff}} &=&\kappa_{0} \Bigg\{1+\frac{1}{2}\frac{ n_{p}^{b}}{n_{0}}\Bigg[\frac{Q^{2}(0)}{e^{2}} \nonumber\\
&&\,-(N_{+}\phi_{+}^{2}(0)+N_{-}\phi_{-}^{2}(0))\Bigg]\Bigg\}^{1/2}.
 \label{eq:kappai02}
\end{eqnarray}
Finally, the Debye-H\"uckel type of equation inside the cavity, $r \leq R$, is given in the form
\begin{eqnarray}
\label{DH1}
\nabla^2\psi &=&  \kappa_{\text{eff}}^2 \psi.
\end{eqnarray}

Within the spherical wall, $R \leq r \leq R+w$, there is no charge and we have the Poisson equation of the form
\begin{eqnarray}
\label{DH2}
\nabla^2\psi &=& 0.
\end{eqnarray}
Outside the enclosure, there are no macroions, only salt ions so for $r \geq R+w$, Eq.~(\ref{PBCR2}) reduces to the linearized DH form
\begin{eqnarray}
\label{DH3}
\nabla^2\psi &=& \kappa_0^2 \psi.
\end{eqnarray}
The boundary conditions Eqs.~(\ref{BC1}) and (\ref{BC2}) can now be written in the form
\begin{eqnarray} 
    \varepsilon_0 \varepsilon_w \frac{\partial\psi}{\partial r}\Big\vert_{r=R} - \varepsilon_0 \varepsilon_m\frac{\partial\psi}{\partial r}\Big\vert_{r=R} &=& \sigma_R \nonumber\\
 \varepsilon_0 \varepsilon_m\frac{\partial\psi}{\partial r}\Big\vert_{r=R+w} - \varepsilon_0 \varepsilon_w \frac{\partial\psi}{\partial r}\Big\vert_{r=R+w} &=& \sigma_{R+w},~~
    \label{BClin}
\end{eqnarray}
where additionally we need to take into account that $\partial \psi/\partial r = 0$ at $r = 0$ and $\partial \psi/\partial r = 0$ for $r \gg R$. 

The last constraint we need to take into account is the number of macroions enclosed inside the nano-reactor, $N_p$. To be consistent with the linearization approximation, we only need the macroion density Eq.~(\ref{nppsi}) to the first order in the electrostatic potential
\begin{eqnarray}
    n_p(\psi) \simeq  n_p^{b} - n_p^{b}Q(0)\beta \psi + {\cal O}\left(\psi^2\right).
\end{eqnarray}
The constraint Eq.~(\ref{const}) then, to the lowest order in $\psi$, assumes the form
\begin{eqnarray}
N_p &=& n_p^{b} V_0 - n_p^{b}Q(0)\beta \int_0^R 4\pi r^2 \psi(r) \text{d}r,
\label{const22}
\end{eqnarray}
where $V_0 = 4\pi R^3/3$ is the volume of the enclosure. 

The solutions of the above linearized PB-CR equations (see Appendix~\ref{appen:A}), together with the boundary conditions and the enclosed macroion number constraint define the full radial profile of the electrostatic potential, the total charge density, and specifically the charge density of the macroions inside the nano-reactor.

Within the linear approximation, the sum of the Maxwell and the van't Hoff components of the pressure tensor yields the pressure inside of the nano-reactor, to the lowest order (quadratic) in $\psi$, as 
\begin{equation}
p_\text{in}
 \simeq \frac{1}{2}\varepsilon_0\varepsilon_w\left[-\left(\frac{\partial\psi}{\partial r}\right)^{2}\Big|_{R}+\kappa_{\text{eff}}^{2}[\psi\left(R\right)]^{2}\right],
 \label{eq:pi01}
\end{equation}
while the pressure outside is
\begin{equation}
p_\text{out}
\simeq \frac{1}{2}\varepsilon_0\varepsilon_w\left[-\left(\frac{\partial\psi}{\partial r}\right)^{2}\Big|_{R+w}+\kappa_{0}^{2}[\psi\left(R+w\right)]^{2}\right],
 \label{eq:po01}
\end{equation}
with $\kappa_{\text{eff}}$ is given by Eq.~(\ref{eq:kappai02}) and the total pressure $p$
is given by Eq.~(\ref{forcetot}).


\section{\label{sec:results}Results}
We now analyze the numerical results obtained in the linearized regime as defined above. The values of the system parameters are $T=300\,\text{K}$, the dielectric constant of the nano-reactor shell $\varepsilon_{m}=5$ and that of the aqueous solution $\varepsilon_{w}=80$, $w=4\,\text{nm}$, and the equilibrium CR constants $n_{0}K_{\pm}=5$, unless otherwise stated. Furthermore, we assume the system can be completely symmetric with $N_+ = N_- = 60$,  partially asymmetric with $N_+ = 60$ and $N_-= 30$, and completely asymmetric with $N_+ = 60$ and $N_-= 0$.


\begin{figure*}[t]
	\centering
	\includegraphics[height=6.44cm]{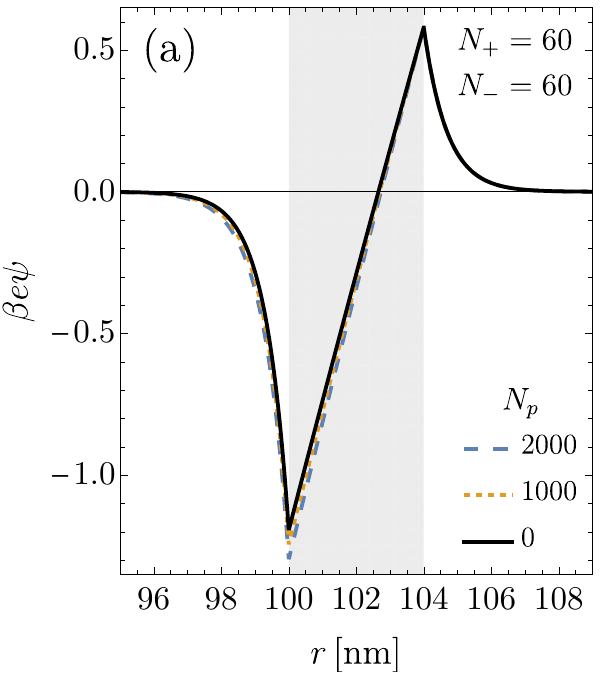}
	\hfill
	\includegraphics[height=6.44cm ,trim={0.9cm 0 0 0},clip]{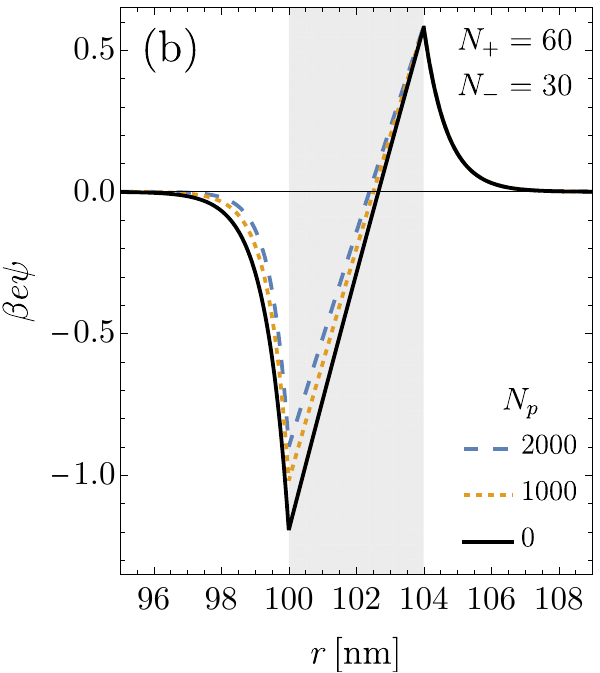}
	\hfill
	\includegraphics[height=6.44cm,trim={0.9cm 0 0 0},clip]{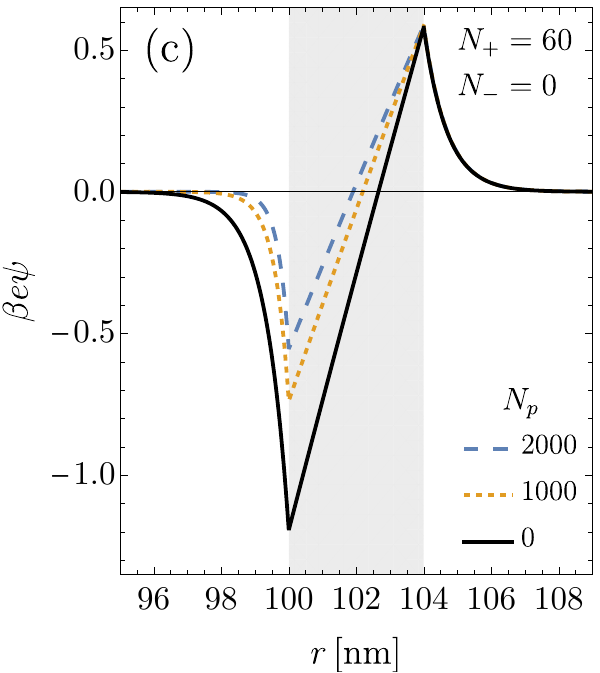}
\caption{$\beta e\psi\left(r\right)$ for different numbers of macroions, $N_{p}$. The plots are evaluated for the completely symmetric case ($N_{+}=60$ and $N_{-}=60$) (a), the partially asymmetric case ($N_{+}=60$ and $N_{-}=30$) (b), and the completely asymmetric case ($N_{+}=60$ and $N_{-}=0$) (c). The nano-reactor shell is indicated by the gray region. The results correspond to $n_{0} = 200\,\text{mM}$, $\sigma_{R} = -0.2\,e/\text{nm}^{2}$, $\sigma_{R + w} = 0.1\,e/\text{nm}^{2}$, $R=100\,\text{nm}$, and
 $w=4\,\text{nm}$.
}
\label{fig2}
\end{figure*}

We first examine the full spatial profile of the rescaled electrostatic potential, $\beta e\psi\left(r\right)$, in the vicinity of the nano-reactor enclosure, obtained from Eqs.~(\ref{DH1})$-$(\ref{DH3}, for different numbers of macroions, $N_{p}=0$, $1000$, and $2000$ as shown in Fig.~\ref{fig2}. The spatial profiles are shown for the completely symmetric case (Fig.~\ref{fig2}(a)), the partially asymmetric case (Fig.~\ref{fig2}(b)), and the completely asymmetric case (Fig.~\ref{fig2}(c)) with $n_{0} = 200\,\text{mM}$, $\sigma_{R} = -0.2\,e/\text{nm}^{2}$, $\sigma_{R + w} = 0.1\,e/\text{nm}^{2}$, and $R=100\,\text{nm}$. Inside the shell, all plots show that the electrostatic potential decreases remarkably in the region close to the inner surface of the nano-reactor, where its magnitude is at a maximum. 
The result also suggests that the electrostatic potential is rather independent of the number of enclosed macroions as shown for the completely symmetric case (Fig.~\ref{fig2}(a)). However, the dependence on the number of entrapped macroions is noticeable for the partially asymmetric case (Fig.~\ref{fig2}(b)), and becomes more appreciable for the completely asymmetric case (Fig.~\ref{fig2}(c)). For all cases, since there are no macroions outside the shell, the electrostatic potentials  outside the nano-reactor are thus identical. As suggested by our results, a higher number of the macroions provides the better electrostatic screening, i.e., the magnitude of electrostatic potential at the inner surface becomes lower, because the nonzero effective charges of the macroions which act as screening charges become higher as shown in Figs.~\ref{fig2}(b) and (c).

\begin{figure*}[t]
	\centering
	\includegraphics[height=6.96cm]{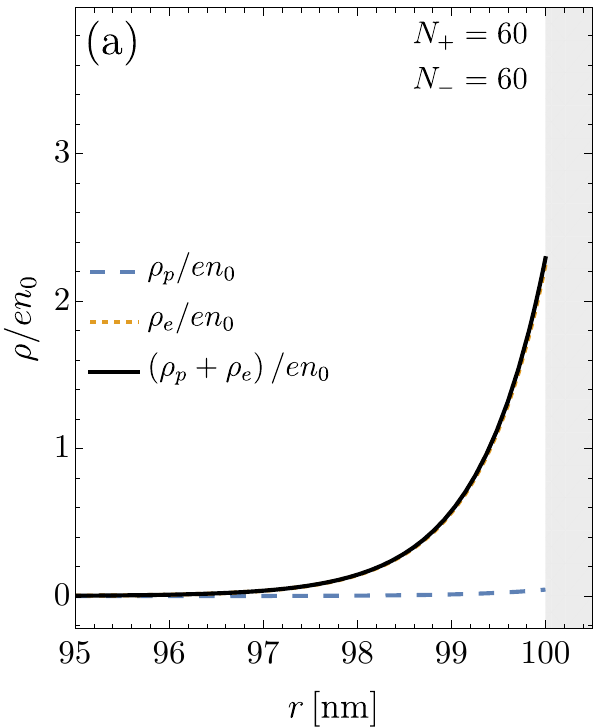}
	\hfill
	\includegraphics[trim={0.9cm 0 0 0},clip,height=6.96cm]{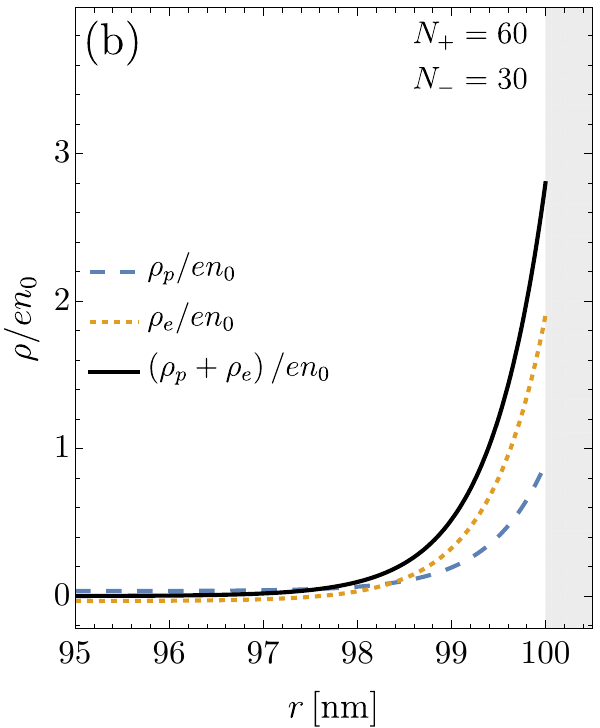}
	\hfill
	\includegraphics[trim={0.9cm 0 0 0},clip,height=6.96cm]{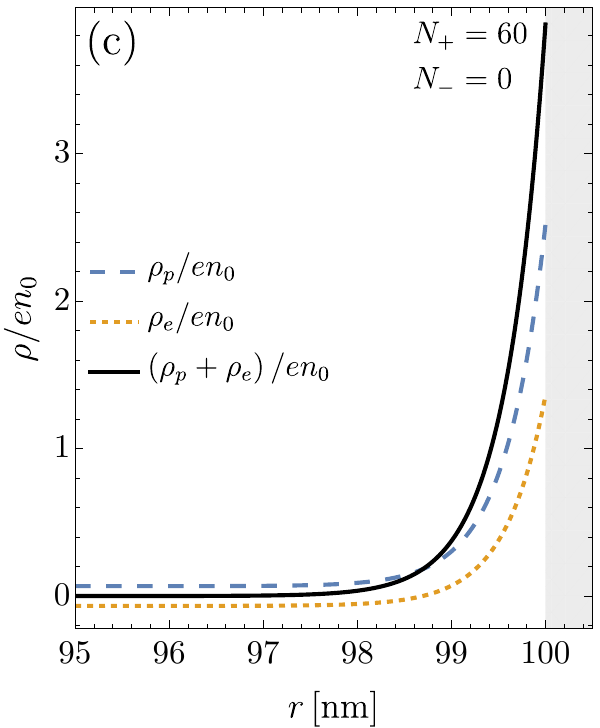}
\caption{The spatial distributions of 
the charge density rescaled by $en_{0}$ for the completely symmetric (a), the partially asymmetric (b), and the completely asymmetric (c) cases. $\rho_{p}$ and $\rho_{e}$ are the charge densities of the macroions and electrolyte ions, respectively. The total number of the macroions is $N_{p}=1000$. $\sigma_{R} = -0.2\,e/\text{nm}^{2}$ and $n_{0} = 200\,\text{mM}$.}
\label{fig4}
\end{figure*}

We next compare the rescaled charge density distribution in Fig.~\ref{fig4} for (a) $(N_+ = N_- = 60)$, (b) $(N_+ = 60, N_- = 30)$, and (c) $(N_+ =60, N_- = 0)$ in the vicinity of the inner surface of the nano-reactor enclosure. For each plot, the charge density of the macroions, $\rho_{p}$, the charge density of the mobile ions, $\rho_{e}$, and the total charge density, 
$\rho_{p}+\rho_{e}$, rescaled by $en_{0}$, are presented. The particular values of the parameters are the same as those in Fig.~\ref{fig2}, with $N_{p}=1000$. In Fig.~\ref{fig4}(a), we can see that the charge density of the macroions inside the nano-reactor is 
nearly flat. 
As a result, the local charges of cations and anions contribute mostly to the total charge density. 
For the partially asymmetric case (Fig.~\ref{fig4}(b)), the contribution of the effective charges of the macroions to the overall charge density becomes significant, resulting in an even higher total charge density in the region close to the inner surface of the nano-reactor enclosure. As the distance far away from the inner surface is increased, the local charge densities of the ions and macroions decrease to a constant, non-vanishing densities with opposite values 
that can be clearly seen in Fig.~\ref{fig4}(c) for the completely asymmetric case. 

For the asymmetrically charged macroions (Figs.~\ref{fig4}(b) and (c)),  the macroions are positively charged not only near the inner surface of the nano-reactor shell but also far away from the surface, 
where the binding rates of $N_{+}$ and $N_{-}$ dissociable surface sites with cations and anions are typically not equal, leaving excess anions in the solution. This result implies that, with increasing distance away from the inner surface, the overall electrolyte ions become negatively charged at a particular point, where the ion density reverses its sign. 
We also note that the total charge density obeys the electro-neutrality condition in the vicinity of the nano-reactor core, where the electrostatic potential vanishes.

\begin{figure*}[t]
	\centering
	\includegraphics[height=6.18cm]{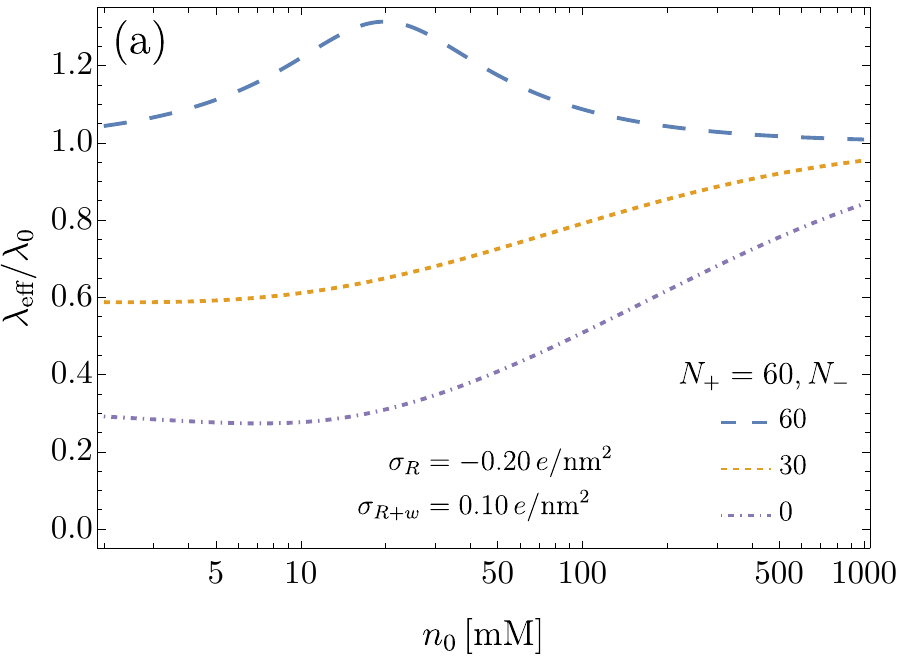}
     \hfill
	\includegraphics[trim={0.85cm 0 0 0},clip,height=6.18cm]{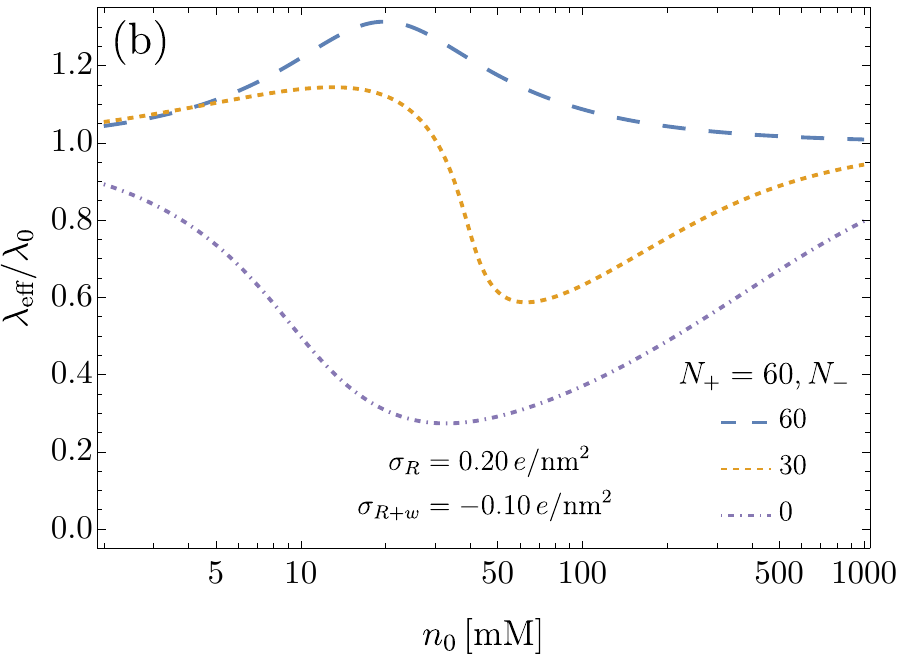}
\caption{The rescaled effective screening length
$\lambda_{\text{eff}}/\lambda_{0}$, where 
$\lambda_{\text{eff}} = \kappa_{\text{eff}}^{-1}$ and 
$\lambda_{0} = \kappa_{0}^{-1}$, as a function of bulk salt concentration $n_{0}$, in the case of (a) $\sigma_{R} = -0.2\,e/\text{nm}^{2}$, $\sigma_{R + w} = 0.1\,e/\text{nm}^{2}$  and (b) $\sigma_{R} = 0.2\,e/\text{nm}^{2}$, $\sigma_{R + w} = -0.1\,e/\text{nm}^{2}$. $N_{p}$ is set as 1000 and $R=100\,\text{nm}$. In each plot, the completely symmetric case ($N_{+} = N_{-} = 60$), the partially asymmetric case ($N_{+} = 60$ and $N_{-} = 30$), and the completely asymmetric case ($N_{+} = 60$ and $N_{-} = 0$) are shown.}
\label{fig6}
\end{figure*}

Next, we examine the dependence of the rescaled effective screening length, $\lambda_{\text{eff}}/\lambda_{0}$, in which $\lambda_{\text{eff}} = \kappa_{\text{eff}}^{-1}$ and 
$\lambda_{0} = \kappa_{0}^{-1}$, on the bulk salt concentration $n_{0}$, inside the nano-reactor enclosure. 
In order to see the effects of the surface charge densities of the nano-reactor on the $\lambda_{\text{eff}}/\lambda_{0}$, the plots are evaluated at the negative inner and positive outer surfaces (Fig.~\ref{fig6}(a)), and at 
the positive inner and negative outer surfaces (Fig.~\ref{fig6}(b)), where the magnitudes of the surface charge densities are $|\sigma_{R}| = 0.2\,e/\text{nm}^{2}$ (inner surface) and  $|\sigma_{R + w}| = 0.1\,e/\text{nm}^{2}$ (outer surface) for both Figs.~\ref{fig6}(a) and (b), while $N_{p}$ is set as 1000. 
For the completely symmetric case $(N_{+} = N_{-} = 60)$, our results suggest that the effective screening length with $\lambda_{\text{eff}}/\lambda_{0} > 1$ exhibits a non-monotonic function of $n_{0}$. As shown for both Figs.~\ref{fig6}(a) and (b) (dashed blue lines), $\lambda_{\text{eff}}/\lambda_{0}$ increases to the maximum value at $n_{0} = 19.82$ mM. Further increasing $n_0$ beyond that value, $\lambda_{\text{eff}}/\lambda_{0}$ decreases monotonically. 
At the partially  asymmetric case ($N_{+} = 60, N_{-} = 30$)
(densely dashed orange curves shown in Figs.~\ref{fig6}(a) and (b)), $\lambda_{\text{eff}}/\lambda_{0}$ monotonically increases as $n_0$ increases with negatively inner and positively outer surfaces (Fig.~\ref{fig6}(a)). However, 
at positively inner and negatively outer surfaces (Fig.~\ref{fig6}(b)), we can observe the different behavior. As $n_{0}$ increases, $\lambda_{\text{eff}}/\lambda_{0}$ increases to the maximum value, followed by a decrease to its minimum at an intermediate $n_{0}$. Beyond that point, $\lambda_{\text{eff}}/\lambda_{0}$ becomes an increasing function of $n_{0}$. The influence of the particular set of surface charge densities $\sigma_{R}$ and $\sigma_{R+w}$ is clearly observed for the completely asymmetric case ($N_{+} = 60, N_{-} = 0$) as well. As we can see from Figs.~\ref{fig6}(a) and (b) (dash-dotted purple curves), as $n_0$ increases, the effective screening length slightly decreases to the minimum value at $n_0$ around 7 mM and, later, highly increases as $n_0$ proceeds
at $\sigma_{R} = - 0.2\,e/\text{nm}^{2}$ and  $\sigma_{R + w} = 0.1\,e/\text{nm}^{2}$ (Fig.~\ref{fig6}(a)). While, at $\sigma_{R} =  0.2\,e/\text{nm}^{2}$ and  $\sigma_{R + w} = - 0.1\,e/\text{nm}^{2}$, 
$\lambda_{\text{eff}}/\lambda_{0}$ exhibits a non-monotonic function of $n_{0}$ in which the screening length remarkably reduces to the minimum at $n_0$ around 33 mM (Fig.~\ref{fig6}(b)). 

\begin{figure*}[t]
	\centering
	\includegraphics[height=6.19cm]{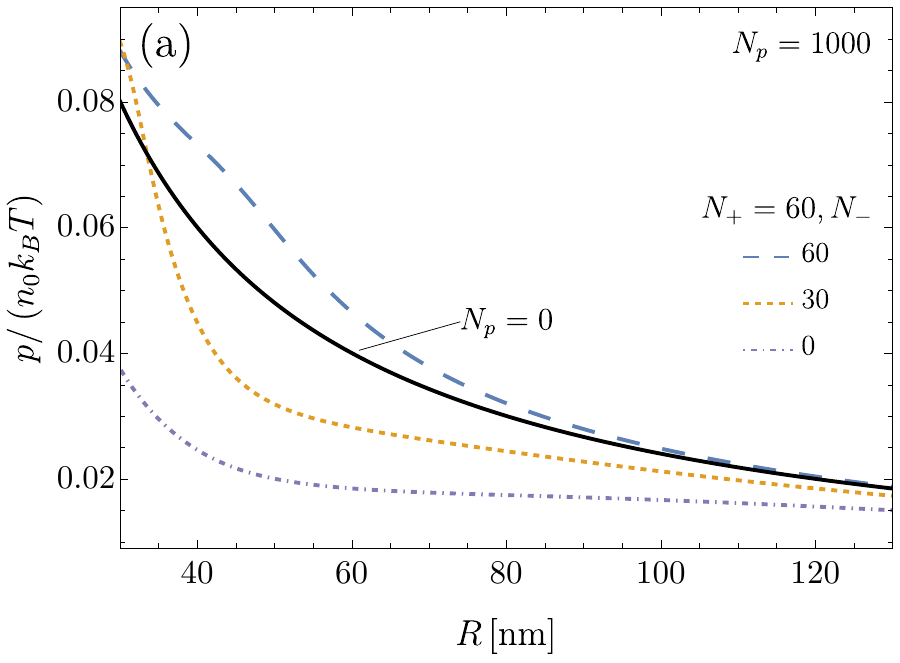}
	\hfill
	\includegraphics[trim={0.9cm 0 0 0},clip,height=6.19cm]{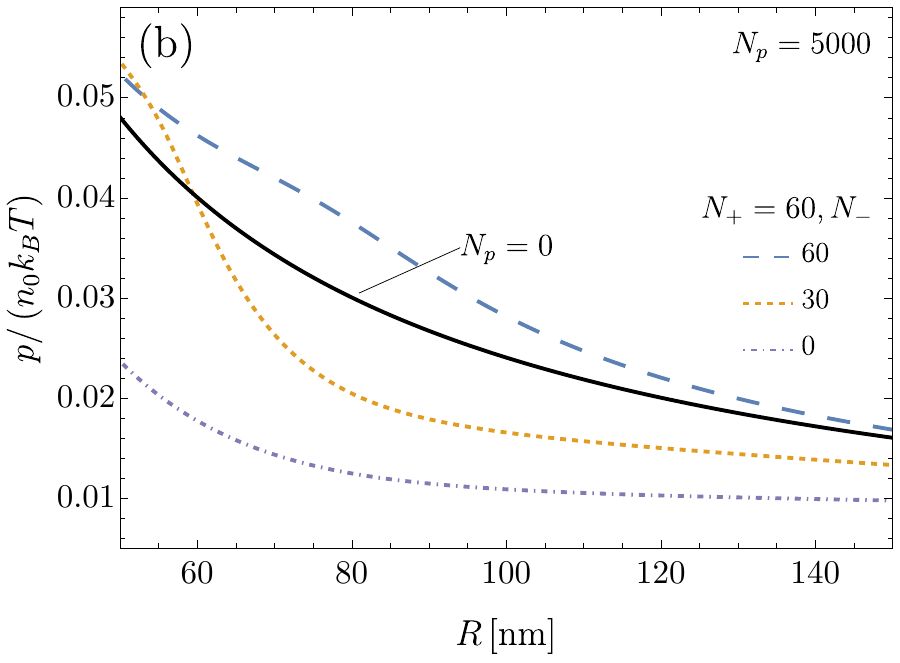}
\caption{The scaled pressure, $p/(n_{0}k_{B}T)$, as a function of the nano-reactor radius, $R$, for $N_{p}=1000$ (a) and $N_{p}=5000$ (b). The solid line represents the pressure in the absence of macroions, $N_{p}=0$. The results correspond to $n_{0} = 200\,\text{mM}$, $\sigma_{R} = -0.2\,e/\text{nm}^{2}$, and $\sigma_{R + w} = 0.1\,e/\text{nm}^{2}$.}
\label{fig7}
\end{figure*}

\begin{figure*}[t]
	\centering
	\includegraphics[height=6.03cm]{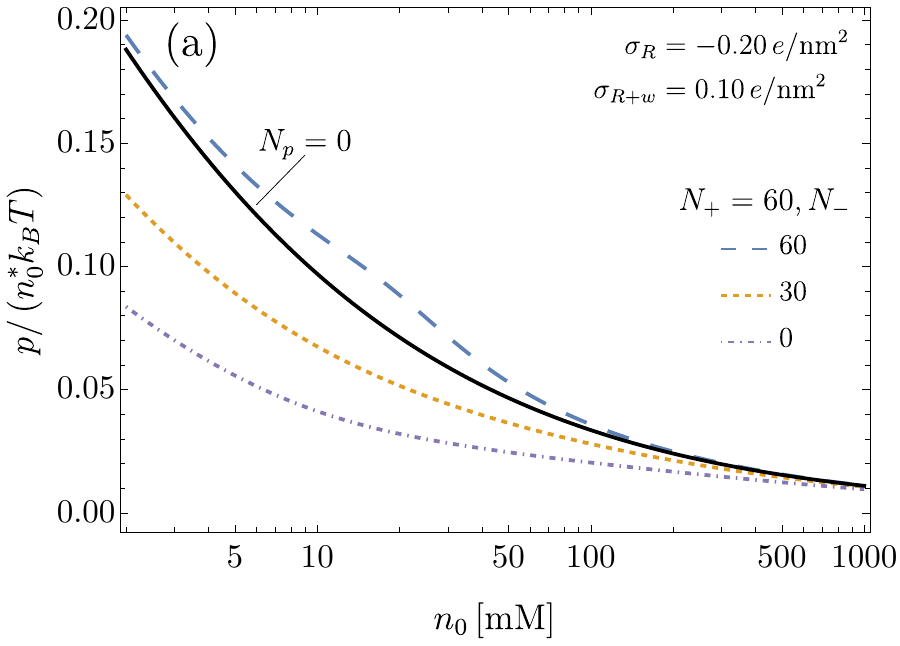}
	\hfill
	\includegraphics[trim={0.9cm 0 0 0},clip,height=6.03cm]{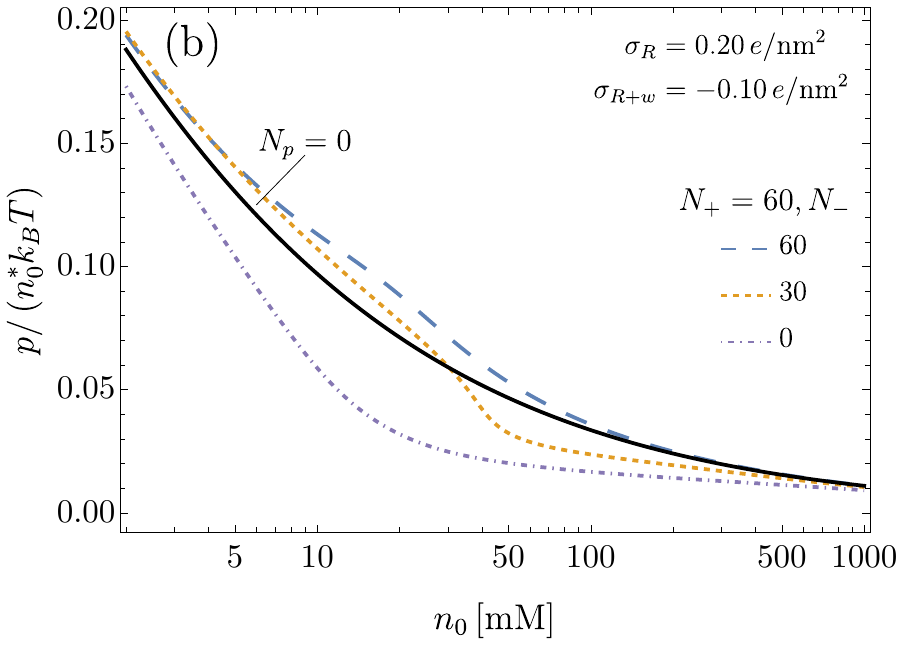}
\caption{The scaled pressure, $p/(n_{0}^{*}k_{B}T)$, with $n_{0}^{*} = 200\,\text{mM}$, as a function of bulk salt concentration, $n_{0}$, for two different sets of surface charge densities: $\sigma_{R} = -0.2\,e/\text{nm}^{2}, \sigma_{R + w} = 0.1\,e/\text{nm}^{2}$ (a) and $\sigma_{R} = 0.2\,e/\text{nm}^{2}, \sigma_{R + w} = -0.1\,e/\text{nm}^{2}$ (b). The results correspond to $R=100\,\text{nm}$, and $N_{p}=1000$. The black solid line represents $N_p = 0$.}
\label{fig8}
\end{figure*}
Finally, we examine the total pressure on the boundary of the nano-reactor enclosure, $p=p_{\text{in}}-p_{\text{out}}$, obtained from Eqs.~(\ref{eq:pi01}) and (\ref{eq:po01}), as a function of the nano-reactor radius $R$ and the bulk salt concentration $n_{0}$. Figure~\ref{fig7} shows the scaled pressure, $p/(n_{0}k_{B}T)$, as a function of $R$ for (a) $N_{p}=1000$ and (b) $N_{p}=5000$. For each plot, the completely symmetric, the partially asymmetric, and the completely asymmetric cases are shown. In addition, the pressure in the absence of macroions, $N_{p}=0$ (black solid line), is also shown for comparison. The other system parameters are $n_{0} = 200\,\text{mM}$, $\sigma_{R} = -0.2\,e/\text{nm}^{2}$, and $\sigma_{R + w} = 0.1\,e/\text{nm}^{2}$. We can see that, while keeping the number of macroions inside the nano-reactor fixed, the net pressure decreases monotonically with increasing nano-reactor radius $R$ for all cases. 
For both Figs.~\ref{fig7}(a) and (b), at the completely symmetric case, the net pressure is higher than that of the case with no macroions inside the nano-reactor. While the pressure for the completely asymmetric case
is lower than that with $N_p = 0$. Interestingly, at small $R$, the pressure for the partially asymmetric case is higher than that 
for the case with no macroions. Further increasing $R$, the pressure decreases below the pressure with $N_p = 0$. The radius $R$ at which the pressure for the partially asymmetric macroions becomes lower than the pressure with $N_p = 0$ clearly depends on the number of macroions.

Figure~\ref{fig8} shows the dependence of the scaled pressure, $p/(n_{0}^{*}k_{B}T)$, with $n_{0}^{*} = 200\,\text{mM}$, on bulk salt concentration $n_{0}$ for surface charge densities (a) $\sigma_{R} = -0.2\,e/\text{nm}^{2}, \sigma_{R + w} = 0.1\,e/\text{nm}^{2}$ and (b) $\sigma_{R} = 0.2\,e/\text{nm}^{2}, \sigma_{R + w} = -0.1\,e/\text{nm}^{2}$. 
Here, the nano-reactor radius and the number of macroions are fixed at $R=100\,\text{nm}$ and $N_{p}=1000$, respectively. The black solid lines represent the results in the absence of macroions. For both Figs.~\ref{fig8}(a) and (b), the results show that the pressure decreases monotonically with increasing the bulk salt concentration 
$n_{0}$, similar to the previous results with increasing the nano-reactor radius $R$ (Fig.~\ref{fig7}). At low  $n_{0}$, more
asymmetric macroions decrease the pressure inside the nano-reactor with negative inner and positive outer surfaces (Fig.~\ref{fig8}(a)). While the difference is hardly seen for the completely symmetric and partially asymmetric macroions with positive inner and negative outer surface charge densities of the enclosure (Fig.~\ref{fig8}(b)). 
At intermediate $n_0$, the pressure for the completely asymmetric macroions is generally lower than that with no macroions, while it becomes higher for the completely symmetric case for both Figs.~\ref{fig8}(a) and (b). However, the difference in the pressure behavior at the partially symmetric macroions with different surface charge densities is observed. As suggested by our results, 
for the positive inner and negative outer surfaces (Fig.~\ref{fig8}(b)), the value of the rescaled pressure is higher than that with $N_p = 0$. As $n_0$ is increased beyond 30 mM, the pressure is dropped below the pressure at $N_p = 0$. The behavior is remarkably different for the negative inner and positive outer surfaces (Fig.~\ref{fig8}(a)) in which the value of the rescaled pressure is always lower than that with $N_p = 0$. 
For high $n_{0}$, the results for both Figs.~\ref{fig8}(a) and (b) clearly suggest that the pressure inside the nano-reactor with various combinations of $(N_{+}, N_{-})$ shows a slightly different from the pressure at the absence of macroions.

The dependence of the pressure on the bulk salt concentration can be understood from $\lambda_{\text{eff}}$ (Eqs.~(\ref{eq:pi01}) and (\ref{eq:po01})). As we can see clearly in the symmetric case, the pressure decreases monotonically as $n_{0}$ is increased with a small peak at $n_{0}$ around 19.50 mM for both Figs.~\ref{fig8}(a) and (b). The small peak of the pressure as $n_0$ increases is slightly different from the peak observed in the plot of $\lambda_{\text{eff}}/\lambda_{0}$ vs $n_0$ shown in Figs.~\ref{fig6}(a) and (b) where the maximum $\lambda_{\text{eff}}/\lambda_{0}$ is found at $n_0$ around 19.82 mM. As we have already seen, the macroions with different charge asymmetries can lead to different pressure variations inside the nano-reactor enclosure, thereby affecting the stability of the system. Likewise, the resulting pressure is strongly influenced by the surface charge densities of the nano-reactor.



\section{\label{sec:cons}Summary and discussion}
In this work, we developed a system model consisting of an electrolyte-permeable nano-reactor immersed in a simple ionic solution. 
Electrolyte ions are exchangeable between the inside and the outside of the spherical nano-reactor enclosure, while charge-regulated macroions are 
confined within the enclosure. 
The key difference in the equilibrated macroions is that the number of macroions inside the nano-reactor enclosure is fixed, depending on the method of experimental preparation. 
By applying appropriate boundary conditions and constraints, we derived the PB-CR equations, focusing on linearized 
Debye-H\"uckel-type approximations. We then analyzed important features of the system model.

In summary, asymmetrically charged macroions with nonzero effective charges enhance electrostatic screening such that a higher number of the macroions provides the better electrostatic screening. While, the electrostatic potential is rather independent of the number of enclosed symmetric macroions with an overall neutral effective charges. The charge regulation equilibria of macroion surface sites are determined by the bathing external solution and effective charges interacting with the background electrostatic potential, characterizing the non-monotonic behavior of the effective screening length, $\lambda_{\textrm{eff}}$. For the negative (positive) inner surface and positive (negative) outer surface, similar non-monotonic behaviors of $\lambda_{\textrm{eff}}$ when $n_0$ increases have been observed
for the completely symmetric macroions. Further investigation shows that the non-monotonicity has also been found for the partially asymmetric macroions at positive inner and negative outer surfaces, and for the completely asymmetric case at negative (positive) inner surface and positive (negative) outer surface. This observation pronounced the strong effect of the particular set of surface charge densities of the nano-reactor on the behavior of $\lambda_{\textrm{eff}}$. 
The total pressure acting on the boundary of the spherical enclosure decreases monotonically with increasing nano-reactor radius and bulk salt concentration. At fixed number of macroions, $N_p$, our results suggest that asymmetric charged macroions can reduce 
the net pressure inside 
the cavity. While, the symmetric macroions always show the pressure which is higher than that at $N_p = 0$. The variation of $N_p$ clearly affects the pressure inside the nano-reactor enclosure, thereby modifying the stability of the confined environment. 
Also, the pressure is strongly determined by the surface charge densities of the enclosure.

As discussed previously, electrostatic interactions and surface charge regulation play a crucial role in determining both the equilibrium properties and stability of nanostructured materials in confined environments. Within dilute bulk and weak electrostatic regimes, the linearized system model provides quantitative results comparable to non-linear analysis. Clearly, the presence of charge regulated macroions modified the screening length and, thus, altered the electrostatic potential. Finally, future analysis may take into account the effect of the correlation between local surface charge sites of the macroions. An extension to a model system with finite dilution of nano-reactors is also of interest.

\section*{Acknowledgements}
We thank Prof. Rudolf Podgornik for many fruitful
discussions, particularly during his visit in Ramkhamhaeng University in 2023-2024. This work is supported by the Zhejiang Key Laboratory of Soft Matter Biomedical Materials (2025ZY01036, 2025E10072). We also acknowledge the funding from the Grant No. 12034019 of the Key Project of the National Natural Science Foundation of China. 

\section{Appendix}\label{sec:appendix}

\subsection{Solutions of Eqs.~(\ref{DH1}), (\ref{DH2}),  and (\ref{DH3})\label{appen:A}}
The fundamental linearized mean-field equations for the electrostatic potential in spherical symmetry, where it depends only on the radial coordinate, can be written in spherical coordinates as variants of the Helmholtz equation
\begin{eqnarray}
\label{DH11}
\frac{\text{d}^2\psi(r)}{\text{d}r^2} + \frac{2}{r} \frac{\text{d}\psi(r)}{\text{d}r} &=&  \kappa_i^2 \psi(r),
\end{eqnarray}
with the solutions in terms of the spherical Bessel functions that can be obtained in the form
\begin{eqnarray}
\psi(r) &=& A \frac{\sinh(\kappa_i r)}{r}  \quad  {\rm for} \quad r \leq R. 
\end{eqnarray}
Inside the spherical wall, the solution of the Poisson equation Eq.~(\ref{DH2}) is obtained as
\begin{eqnarray}
\psi(r)= \frac{B}{r} + C, \quad  {\rm for} \quad R\leq r\leq R+w,
\end{eqnarray}
and finally, outside the nano-reactor, the solution of 
Eq.~(\ref{DH3}) is
\begin{eqnarray}
\psi(r) = D~ \frac{{\rm e}^{-\kappa_0 r}}{r}  \quad {\rm for} \quad r \geq R+w.
\end{eqnarray}
This is the fundamental set of solutions that now needs to be combined with the boundary conditions Eq.~(\ref{BClin}), yielding the complete form of the electrostatic potential. The constants $A$, $B$, $C$, and $D$ are
\begin{eqnarray*}
A &=& R\Bigg\{2 \ell_{R+w} R \left[\varepsilon_{m} R + \varepsilon_{w} w + 
      \varepsilon_{w} \kappa_{0} w (R + w)\right] \textrm{sgn}(\sigma_{R}) \nonumber\\
     && \quad +\frac{
   2 \varepsilon_{m} \ell_{R} (R + w)^{2} }{\textrm{sgn}(\sigma_{R+w})}\Bigg\}/\Delta,\\
B &=& 2 \varepsilon_{w} R (R + w) \Bigg\{\frac{\ell_{R+w} R \left[1 + \kappa_{0} (R + w)\right] \sinh(\kappa_{i} R)}{\textrm{sgn}(\sigma_{R})} \nonumber\\
   && \quad +\frac{
   \ell_{R} (R + w) \left[\sinh(\kappa_{i} R)-\kappa_{i} R \cosh(\kappa_{i} R)\right]}{  \textrm{sgn}(\sigma_{R+w})}\Bigg\}/\Delta,\\
C &=&\{ 2 \ell_{R} (R + w)^{2}\textrm{sgn}(
   \sigma_{R+w}) \nonumber\\
   && \quad \times \left[\varepsilon_{w} \kappa_{i} R \cosh(
      \kappa_{i} R) + (\varepsilon_{m} - \varepsilon_{w}) \sinh(\kappa_{i} R)\right] \nonumber\\
   &&\quad -2 \ell_{R+w} R^{2}\left[\varepsilon_{w} -\varepsilon_{m} + \varepsilon_{w} \kappa_{0} (R + w)\right] \nonumber\\
   &&\quad \times \textrm{sgn}(
   \sigma_{R})\sinh(\kappa_{i} R) \}/\Delta,\\
D &=& e^{\kappa_{0} (R + w)} (R + 
   w) \{2 \varepsilon_{m} \ell_{R+w} R^{2} \textrm{sgn}(\sigma_{R}) \sinh(\kappa_{i} R) \nonumber\\
   &&\quad + 2\ell_{R} (R + w)\textrm{sgn}(\sigma_{R+w})[\varepsilon_{w} \kappa_{i} R w \cosh(
        \kappa_{i} R) \nonumber\\
        &&\quad        
        + (\varepsilon_{m} (R + w) -\varepsilon_{w} w )\sinh(\kappa_{i} R)]\}/\Delta,
\end{eqnarray*}
where 
\begin{eqnarray*}
\Delta &=& \ell_{R} \ell_{R+w}\{ \kappa_{i} R \left[\varepsilon_{m} R + \varepsilon_{w} w + 
      \varepsilon_{w} \kappa_{0} w (R + w)\right] \cosh(
     \kappa_{i} R) \nonumber\\
    &&\quad + \left[(\varepsilon_{m} - \varepsilon_{w}) w + 
      \kappa_{0} (R + w) (\varepsilon_{m} (R + w) -\varepsilon_{w} w )\right]\nonumber\\ &&\quad \times \sinh(\kappa_{i} R)\}.
\end{eqnarray*}
$\ell_{R}$ and $\ell_{R + w}$ are the Gouy-Chapman lengths defined by $ \ell_{R} = 2\varepsilon_{0}\varepsilon_{w}k_{\text{B}}T/(e|\sigma_{R}|)$ and $\ell_{R + w} = 2\varepsilon_{0}\varepsilon_{w}k_{\text{B}}T/(e|\sigma_{R + w}|)$.

\bibliographystyle{apsrev4-2}
\bibliography{nanoreactors}

\end{document}